\def\year{2021}\relax
\documentclass[letterpaper]{article} 
\usepackage{aaai21}  
\usepackage{times}  
\usepackage{helvet} 
\usepackage{courier}  
\usepackage[hyphens]{url}  
\usepackage{graphicx} 
\usepackage{natbib}  
\usepackage{caption} 
\usepackage{times}
\usepackage{soul}
\usepackage{url}
\usepackage{graphicx}
\usepackage{amsmath}
\usepackage{amsthm}
\usepackage{amssymb}
\usepackage{booktabs}
\usepackage{algorithm}
\usepackage{algorithmic}
\usepackage{comment}
\usepackage{subfigure}
\usepackage{multirow}
\usepackage{multicol}  
\usepackage{enumitem}
\usepackage{array}

\title{DEAR: Deep Reinforcement Learning for Online Advertising Impression\\in Recommender Systems}
\author{
	Xiangyu Zhao\textsuperscript{\rm 1},
	Changsheng Gu\textsuperscript{\rm 2}, Haoshenglun Zhang\textsuperscript{\rm 2}\\
	Xiwang Yang\textsuperscript{\rm 2}, Xiaobing Liu\textsuperscript{\rm 2}, Jiliang Tang\textsuperscript{\rm 1}, Hui Liu\textsuperscript{\rm 1}\\
}
\affiliations{
	\textsuperscript{\rm 1}Michigan State University, \textsuperscript{\rm 2}Bytedance\\
	\{zhaoxi35,tangjili,liuhui7\}@msu.edu\\ \{guchangsheng,zhanghaoshenglun,yangxiwang,will.liu\}@bytedance.com
}
\begin{document}
\maketitle

\begin{abstract}
With the recent prevalence of Reinforcement Learning (RL), there have been tremendous interests in utilizing RL for online advertising in recommendation platforms (e.g., e-commerce and news feed sites). However, most RL-based advertising algorithms focus on optimizing ads' revenue while ignoring the possible negative influence of ads on user experience of recommended items (products, articles and videos). Developing an optimal advertising algorithm in recommendations faces immense challenges because interpolating ads improperly or too frequently may decrease user experience, while interpolating fewer ads will reduce the advertising revenue. Thus, in this paper, we propose a novel advertising strategy for the rec/ads trade-off. To be specific, we develop an RL-based framework that can continuously update its advertising strategies and maximize reward in the long run. Given a recommendation list, we design a novel Deep Q-network architecture that can determine three internally related tasks jointly, i.e., (i) whether to interpolate an ad or not in the recommendation list, and if yes, (ii) the optimal ad and (iii) the optimal location to interpolate. The experimental results based on real-world data demonstrate the effectiveness of the proposed framework. 
\end{abstract}

\section{Introduction}
\label{sec:introduction}
Online advertising is a form of advertising that leverages the Internet to deliver promotional marketing messages to consumers. The goal of online advertising is to assign the right ads to the right consumers so as to maximize the revenue, click-through rate (CTR) or return on investment~(ROI) of the advertising campaign. The two main marketing strategies in online advertising are guaranteed delivery~(GD) and real-time bidding~(RTB). For guaranteed delivery, ad exposures to consumers are guaranteed by contracts signed between advertisers and publishers in advance~\cite{Jia2016Efficient}. For real-time bidding, each ad impression is bid by advertisers in real-time when an impression is just generated from a consumer visit~\cite{cai2017real}. However, the majority of online advertising techniques are based on offline/static optimization algorithms that treat each impression independently and maximize the immediate revenue for each impression, which is challenging in real-world business, especially when the environment is unstable. Therefore, great efforts have been made on developing reinforcement learning-based online advertising techniques~\cite{cai2017real,wang2018learning,rohde2018recogym,wu2018budget,jin2018real}, which can continuously update their advertising strategies during the interactions with consumers, and the optimal strategy is made by maximizing the expected long-term cumulative revenue from consumers. However, most existing works focus on maximizing the income of ads, while ignoring the negative influence of ads on user experience for recommendations.

\begin{figure}
	\centering
	\includegraphics[width=\linewidth]{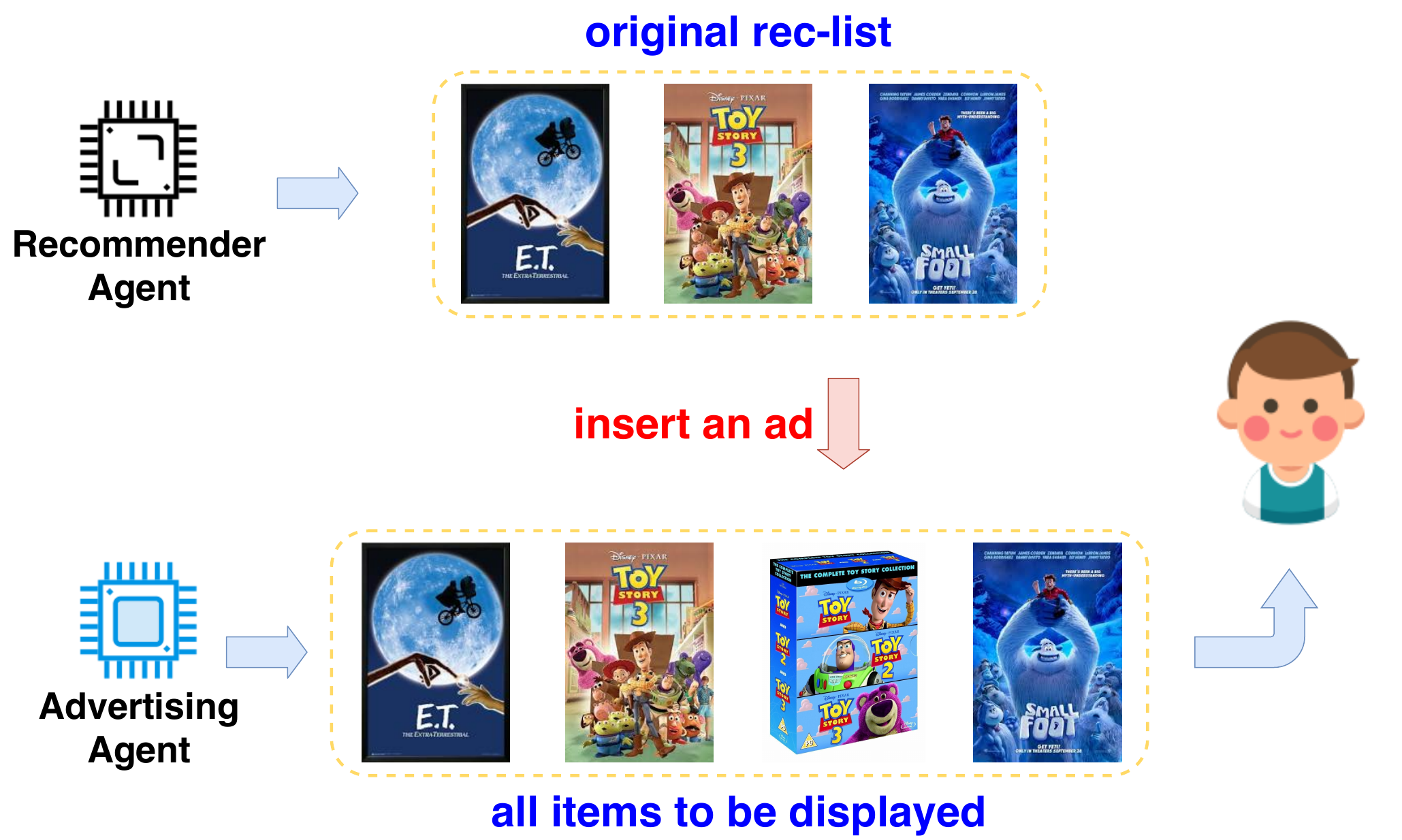}
	\caption{An example of online advertising impression}
	\label{fig:Fig1_example}
\end{figure}

Designing an appropriate advertising strategy is a challenging problem, since (i) displaying too many ads or improper ads will degrade user experience and engagement; and (ii) displaying insufficient ads will reduce the advertising revenue of the platforms. In real-world platforms, as shown in Figure~\ref{fig:Fig1_example}, ads are often displayed with normal recommended items, where recommendation and advertising strategies are typically developed by different departments, and optimized by different techniques with different metrics~\cite{feng2018learning}. Upon a user's request, the recommendation system generates a list of recommendations according to user's interests. Then, the advertising system needs to make three decisions (sub-actions), i.e., whether to interpolate an ad in current \textit{recommendation list} (rec-list); and if yes, the advertising system also needs to choose the optimal ad and interpolate it into the optimal location (e.g., in Figure~\ref{fig:Fig1_example} the \textit{advertising agent} (AA) interpolates an ``ad of toy'' after the second movie of the rec-list). The first sub-action maintains the frequency of ads, while the other two sub-actions aim to control the appropriateness of ads. The goal of advertising strategy is to simultaneously maximize the income of ads and minimize the negative influence of ads on user experience.

\begin{figure}
	\centering
	\includegraphics[width=\linewidth]{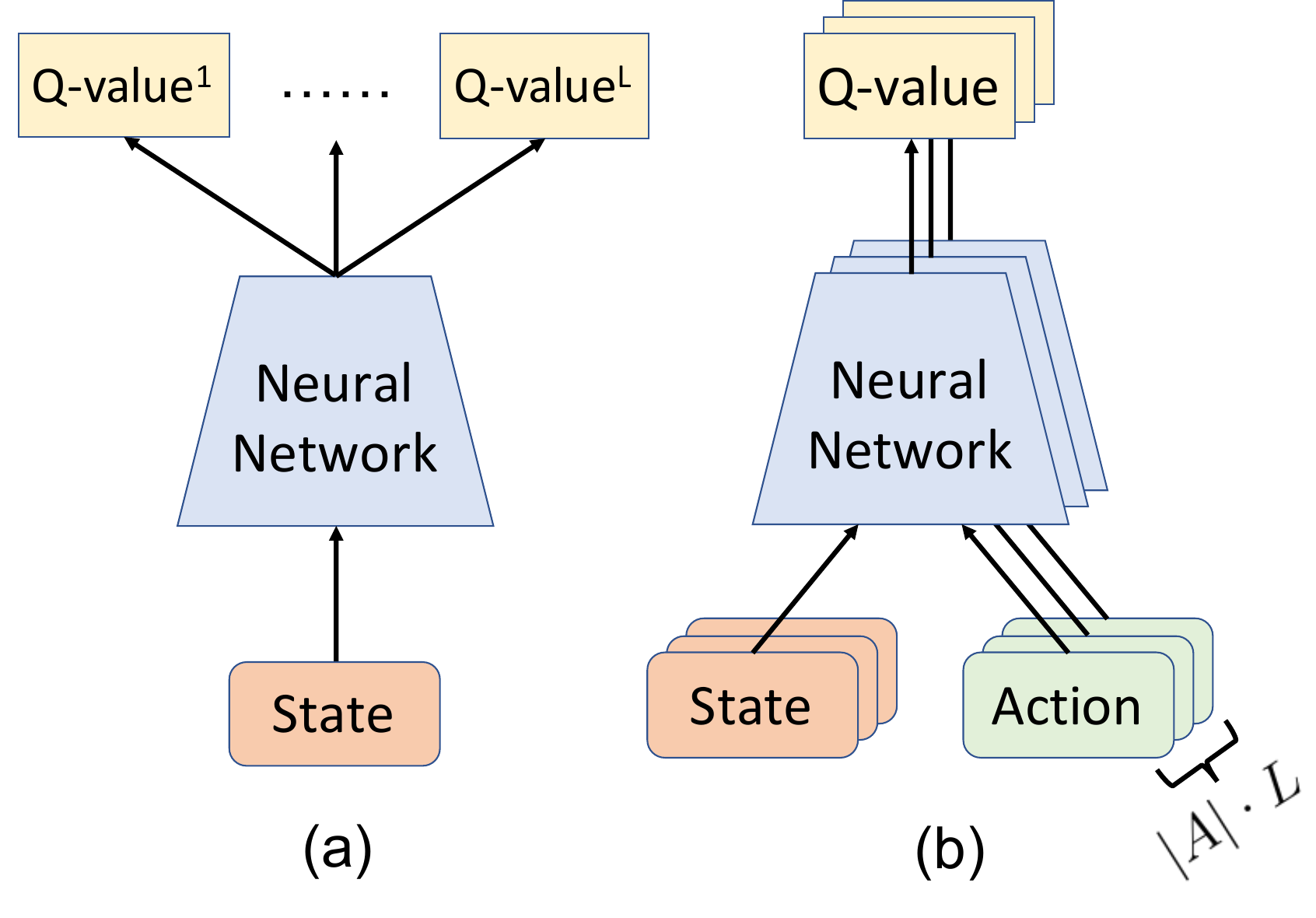}
	\caption{Classic DQN architectures for online advertising.}
	\label{fig:Fig2_DQNs}
\end{figure}

The above-mentioned three decisions (sub-actions) are internally related, i.e., (only) when the AA decides to interpolate an ad, the locations and candidate ads together determine the rewards. Figure~\ref{fig:Fig2_DQNs} illustrates the two conventional Deep Q-network (DQN) architectures for online advertising. Note that in this paper, we suppose (i) there are $|A|$ candidate ads for each request, and (ii) the length of the recommendation list (or rec-list) is $L$. The DQN in Figure~\ref{fig:Fig2_DQNs}(a) takes the state space and outputs Q-values of all locations. This architecture can determine the optimal location but cannot choose the specific ad to interpolate. The DQN in Figure~\ref{fig:Fig2_DQNs}(b) inputs a state-action pair and outputs the Q-value corresponding to a specific action (ad). This architecture can select a specific ad but cannot decide the optimal location. Taking a representation of location (e.g., one-hot vector) as the additional input is an alternative way, but $O(|A|\cdot L)$ evaluations are necessary to find the optimal action-value function $Q^*(s, a)$, which prevents the  DQN architecture from being adopted in practical advertising systems. It is worth noting that both architectures cannot determine whether to interpolate an ad (or not) into a given rec-list. Thus, in this paper, we design a new \textbf{DE}ep reinforcement learning framework with a novel DQN architecture for online \textbf{A}dvertising in \textbf{R}ecommender systems (\textbf{DEAR}), which can determine the aforementioned three tasks simultaneously with reasonable time complexity. We summarize our major contributions as follows:  
\begin{itemize}[leftmargin=*]
	\item We decompose online advertising with recommendations into three related decisions and provide a principled approach to model it;
	\item We propose a deep reinforcement learning-based framework DEAR and a novel Q-network architecture, which can simultaneously determine whether to interpolate an ad, the optimal location and which ad to interpolate;
	\item We demonstrate the effectiveness of the proposed framework in real-world short video site.
\end{itemize}
\section{Problem Statement}\label{sec:problem} 
In this paper, we study the advertising problem within a rec-list as a Markov Decision Process (MDP), in which an Advertising-Agent (AA) interacts with environment $\mathcal{E}$ (or users) by sequentially interpolating ads into a sequence of rec-lists over time, so as to maximize the cumulative reward from the environment. Formally, the MDP consists of a tuple of five elements $(\mathcal{S}, \mathcal{A}, \mathcal{P}, \mathcal{R}, \gamma)$:
\begin{itemize}[leftmargin=*]
	\item {\bf State space $\mathcal{S}$}: A state $s_t \in \mathcal{S}$ is defined as a user's browsing history before time $t$ and the information of current request at time $t$. More specifically, a state $s_t$ consists of a user's recommendation and ad browsing history, the rec-list, and contextual information of the current request. 
	\item {\bf Action space $\mathcal{A}$}:  The action $a_t = (a_t^{ad}, a_t^{loc}) \in \mathcal{A}$ of AA is to determine three internally related tasks, i.e., whether interpolate an ad in current rec-list (that is involved in $a_t^{loc}$, more details are presented in following sections); if yes, the AA needs to choose a specific ad $a_t^{ad*}$ and interpolate it into the optimal location $a_t^{loc*}$ in the rec-list. Without the loss of generality, we assume that the AA could interpolate at most one ad into a rec-list, but it is straightforward to extend it with multiple ads.
	\item {\bf Reward $\mathcal{R}$}: After the AA taking action $a_t$ at the state $s_t$, i.e., (not) interpolating an ad into a rec-list, a user browses this mixed rec-ad list and provides her feedback. The AA will receive the immediate reward $r(s_t,a_t)$ based on user's feedback. The reward $r(s_t,a_t)$ is two-fold: (i) the income of an ad that depends on the quality of the ad, and (ii) the influence of an ad on the user experience.
	\item {\bf Transition probability $\mathcal{P}$}: Transition probability $p(s_{t+1}|s_t,a_t)$ defines the state transition from $s_t$ to $s_{t+1}$ after taking action $a_t$. We assume that the MDP satisfies $p(s_{t+1}|s_t,a_t,...,s_1,a_1) = p(s_{t+1}|s_t,a_t)$.
	\item {\bf Discount factor $\gamma$}: Discount factor $\gamma \in [0,1]$ is introduced to measure the present value of future reward. When $\gamma = 1$, all future rewards will be fully counted into current action; on the contrary, when $\gamma = 0$, only the immediate reward will be considered.
\end{itemize}

With the above-mentioned notations and definitions, the problem of ad interpolation into recommendation lists can be formally defined as follows: \textit{Given the historical MDP, i.e., $(\mathcal{S}, \mathcal{A}, \mathcal{P}, \mathcal{R}, \gamma)$, the goal is to find an advertising policy $\pi:\mathcal{S} \rightarrow \mathcal{A}$, which can maximize the cumulative reward from users}, i.e., maximizing the income of ads and minimizing the negative influence on user experience. 
\section{The Proposed Framework}
\label{sec:framework}
In this section, we will propose a deep reinforcement learning framework for online advertising in recommender systems. To be more specific, we will first propose a novel DQN architecture, which could simultaneously tackle the aforementioned three tasks. Then, we discuss how to train the framework via offline users' behavior logs. 

\subsection{The DQN Architecture for Online Advertising}
\label{sec:architecture}
As aforementioned, the online advertising in recommender system problem is challenging because (i) the action of the advertising agent (AA) is complex, which consists of three sub-actions, i.e., whether interpolate an ad into the current rec-list, if yes, which ad is optimal and where is the best location; (ii) the three sub-actions are internally related, i.e., when the AA decides to interpolate an ad, the candidate ads and locations are interactive to maximize the reward, which prevents traditional DQN architectures from being employed in online advertising systems; and (iii) the AA should simultaneously maximize the income of ads and minimize the negative influence of ads on user experience. To address these challenges, we propose a deep reinforcement learning framework with a novel Deep Q-network architecture. In the following, we first introduce the processing of state and action features, and then we illustrate the proposed DQN architecture with an optimization algorithm.

\subsubsection{The Processing of State and Action Features}
\label{sec:features}
The state $s_t$ consists of a user's rec/ads browsing history, the contextual information and rec-list of current request. The recommendation (or ad) browsing history is a sequence of recommendations (or ads) the user has browsed. We leverage two RNNs with Gated Recurrent Units (GRU) to capture users' sequential preference of recommendations and ads separately. The inputs of RNN are the features of user's recently browsed recommendations (or ads), while we use the final hidden state of RNN as the representation of user's dynamic preference of recommendations $p_t^{rec}$ (or ads $p_t^{ad}$). Here we leverage GRU rather than Long Short-Term Memory (LSTM) because of GRU's simpler architecture and fewer parameters.

The contextual information feature $c_t$ of current user request consists of information such as the OS (ios or android), app version and feed type (swiping up/down the screen) of user's current request. Next, we represent the rec-list of current request by the concatenated features of $L$ recommended items that will be displayed in current request, and we transform them into a low-dimensional dense vector $rec_t = tanh(W_{rec} concat(rec_1,\cdots,rec_L)+b_{rec})$. Note that other architectures like CNN for NLP~\cite{Kim2014Convolutional} can also be leveraged. Finally, we get a low-dimensional representation of state $s_t$ by concatenating $p_t^{rec}, p_t^{ad}, c_t$ and $rec_t$:
\begin{equation}\label{equ:s_t}
	s_t=concat(p_t^{rec}, p_t^{ad}, c_t, rec_t)
\end{equation}
For the transition from $s_t$ to $s_{t+1}$, the recommendations and ads browsed at time $t$ will be added into browsing history to generate $p_{t+1}^{rec}$ and $p_{t+1}^{ad}$, $c_{t+1}$ depends on user's behavior at time $t+1$, and $rec_{t+1}$ comes from the recommendation system. For the action $a_t = (a_t^{ad}, a_t^{loc}) \in \mathcal{A}$, $a_t^{ad}$ is the feature of a candidate ad, and $a_t^{loc}\in \mathbb{R}^{L+1}$ is the location to interpolate the selected ad (given a list of $L$ recommendations, there exist $L+1$ possible locations). Next, we will elaborate the architecture of the proposed DQN architecture.

\begin{figure}
	\centering
	\includegraphics[width=\linewidth]{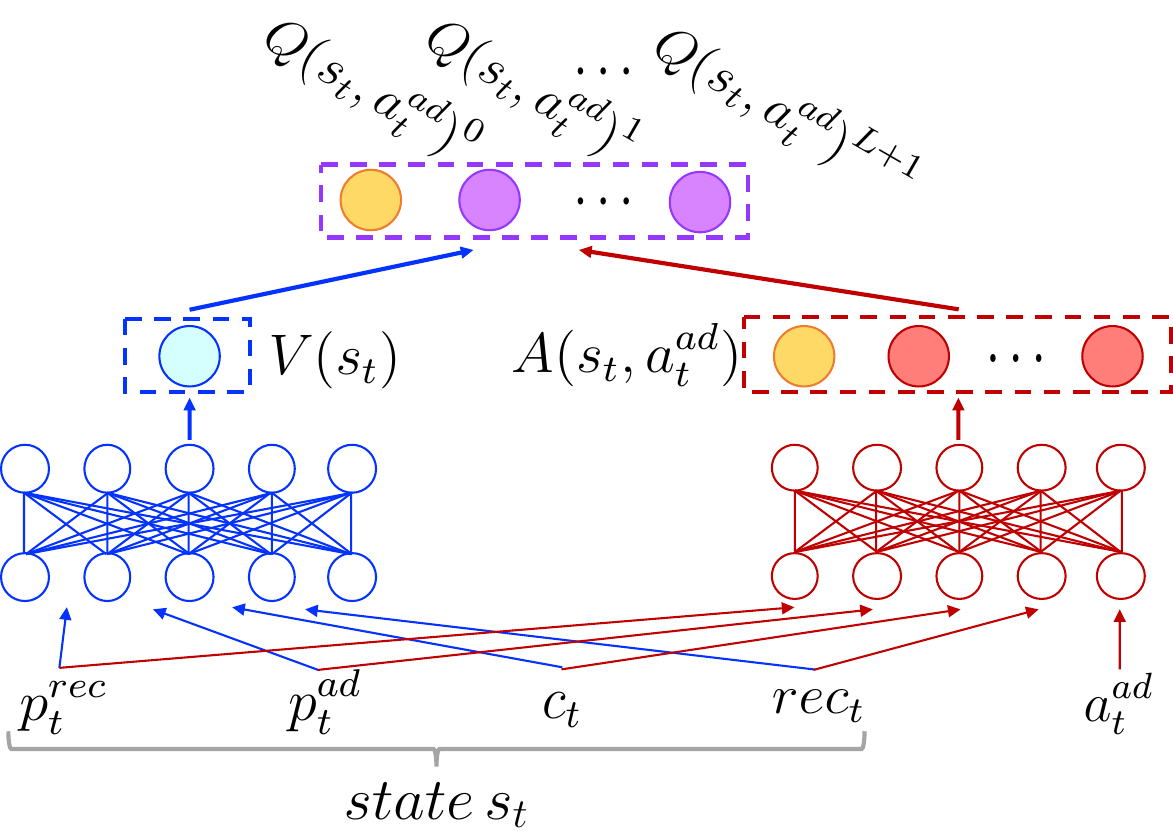}
	\caption{The detailed architecture of the proposed DQN.}
	\label{fig:Fig3_DEAR1}
\end{figure}

\subsubsection{The Proposed DQN Architecture}
\label{sec:DQN}
Given the state $s_t$, the action $a_t$ of AA consists of three sub-actions, i.e., whether to interpolate an ad, if yes, (ii) where is the optimal location and (iii) which ad is optimal.  

\begin{figure}[t]
	\centering
	\includegraphics[width=52mm]{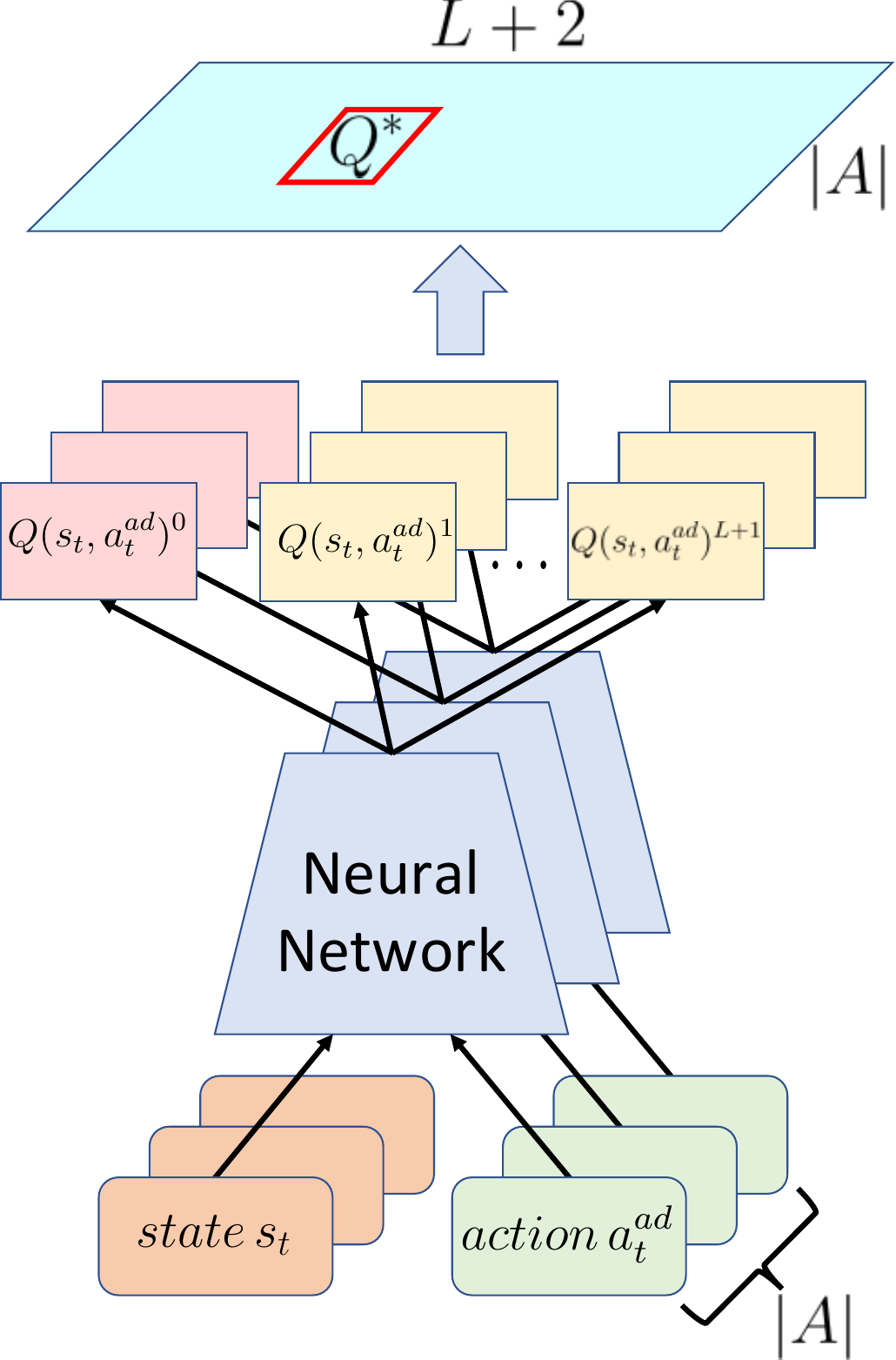}
	\caption{The novel DQN architecture for online advertising.}
	\label{fig:Fig3_DEAR2}
\end{figure}

We first consider simultaneously tackle the sub-action (ii) and (iii). In other words, we aim to estimate the Q-values of all possible locations $a_t^{loc}$ for any given candidate ad $a_t^{ad}$. To incorporate these two sub-actions into one framework, we proposed a novel DQN architecture, as illustrated in Figure~\ref{fig:Fig3_DEAR2}, which is on the top of the two conventional Deep Q-network architectures shown in Figure~\ref{fig:Fig2_DQNs}. The inputs are the representations of state $s_t$ and any candidate ad $a_t^{ad}$, while the output is the action-value (Q-value) corresponding to $L+1$ locations. In this way, the proposed DQN architecture could take advantage of both traditional DQN architectures, which could simultaneously evaluate the Q-values of two types of internally related sub-actions, i.e., evaluating the Q-values of all possible locations for an ad. 

To incorporate the first sub-action (whether to interpolate an ad or not) into the above DQN architecture, we consider not interpolating an ad as a special $location\,0$, and extend the length of output layer from $L+1$ to $L+2$, where $Q(s_t,a_t^{ad})^0$ corresponds to the Q-value of not incorporating an ad into current rec-list. Therefore, the proposed DQN architecture could take the three sub-actions simultaneously, where the Q-value depends on the combination of ad-location pair; and when $Q(s_t,a_t^{ad})^0$ of any candidate ads corresponds to the maximal Q-value, the AA will not interpolate an ad into current rec-list. It worth noting that, compared with the classical DQN in Figure~\ref{fig:Fig2_DQNs} (b) that requires $|A|\cdot L$ times of value-function evaluations forward through the whole neural network, our proposed model reduces the temporal complexity of forward propagations from $O(|A|\cdot L)$ to $O(|A|)$.

The detailed DQN architecture is illustrated in Figure~\ref{fig:Fig3_DEAR1}. On one hand, whether to interpolate an ad into current rec-list is mainly impacted by the state $s_t$ (the browsing history, the contextual information and especially the quality of current rec-list), e.g., if a user has good experience for current rec-list, the advertising agent may prefer to interpolate an ad into the current rec-list; while if a user has bad experience for current rec-list, the user has high possibility to leave, then the AA will not insert an ad to increase this possibility. On the other hand, the reward for choosing an ad and location is closely related to all the features (both current rec-list and the ads). According to this observation, we divide the Q-function into value function $V(s_t)$, which is determined by the state features, and the advantage function $A(s_t, a_t)$, which is determined by the features of both state and action~\cite{Wang2015Dueling}.

\subsubsection{Discussion} 
There exist two classic DQN architectures as illustrated in Figure~\ref{fig:Fig2_DQNs}, where (i) the left one takes state as input and outputs Q-values of all action candidates, and (ii) the right one takes a state-action pair and outputs the Q-value of this pair. Typically, these two conventional architectures can only evaluate Q-values for one level of actions, e.g., the agent in Maze environment can only choose to go up, down, left, or right~\cite{brockman2016openai}. Compared with these two traditional architectures, the proposed DEAR takes a state-action pair of one level of actions, and outputs the Q-values corresponding to the combination of this state-action pair and another level of actions. Hierarchical reinforcement learning (HRL) architectures like~\cite{kulkarni2016hierarchical} can also handle multiple levels of tasks. However, HRL frameworks suffer from the instability problem when training multiple levels jointly~\cite{nachum2018data}. To the best of our knowledge, the proposed DEAR architecture is the \textbf{first individual DQN architecture} that can evaluate the Q-values of multiple levels of internally related actions simultaneously with reasonable time complexity, i.e., our model decreases the temporal complexity of forward propagations from $O(|A|\cdot L)$ to $O(|A|)$.  This design is general which has many other possible applications. For example, in Maze environment the input of DEAR can be the pair of agent's location (state) and the direction to go (action), then the DEAR can output the Q-values corresponding to the location, direction and how many steps to go in this direction (another level of related actions).

\subsection{The Reward Function}
\label{sec:reward}
After the AA executing an action $a_t$ at the state $s_t$, i.e., interpolating an ad into a rec-list (or not), a user browses this mixed rec-ad list and provides her feedback. In online advertising with normal recommendations, the AA aims to simultaneously maximize the income of ads and minimize the negative influence of ads on user experience (i.e., to optimize user experience). Thus the immediate reward $r_t(s_t,a_t)$ is two-fold: (i) the income of ad $r_t^{ad}$, and (ii) the user experience $r_t^{ex}$. 

In practical platforms, the major risk of interpolating ads improperly or too frequently is that the user will leave the platforms. Thus, user experience is measured by whether she/he will leave the platform after browsing the current rec-ad list, and we have: 
\begin{equation}\label{equ:reward2}
r_t^{ex}=
\left\{\begin{array}{ll}
1
&continue\\
-1
& leave
\end{array}\right.
\end{equation}
\noindent In other words, the AA will receive a positive reward if the user continues to browse the next list, otherwise negative reward. Then, we design the reward function as follows:
\begin{equation}\label{equ:reward}
r_t(s_t,a_t)=r_t^{ad} + \alpha \cdot r_t^{ex}
\end{equation}
\noindent where the $r_t^{ad}$ is the income of ad, which is a positive value if interpolating an ad, otherwise 0. The hyper-parameter $\alpha$ controls the importance of the second term, which measures the influence of an ad on user experience. Based on the reward function, the optimal action-value function $Q^*(s_t, a_t)$, which has the maximum expected return achievable by the optimal policy, should follow the Bellman equation~\cite{bellman2013dynamic}:
\begin{equation}\label{equ:Q*sa}
Q^{*}(s_t, a_t)=\mathbb{E}_{s_{t+1}} \, \big[r_t{+}\gamma\max_{a_{t+1}}Q^{*}(s_{t+1}, a_{t+1})|s_t,a_t\big]
\end{equation}
\noindent where the operation $\max_{a_{t+1}}$ needs to look through all candidate ads $\{a_{t+1}^{ad}\}$ (input) and all locations $\{a_{t+1}^{loc}\}$ (output), including the location that represents not inserting an ad.

\begin{algorithm}[t]
	\small
	\caption{\label{alg:model1} Off-policy Training of DEAR Framework.}
	\raggedright
	\begin{algorithmic} [1]
		\STATE Initialize the capacity of replay buffer $\mathcal{D}$
		\STATE Initialize action-value function $Q$ with random weights
		\FOR{session $=1, M$}
		\STATE  Initialize state $s_{0}$ from previous sessions
		\FOR{$t=1, T$}
		\STATE  Observe state $s_t=concat(p_t^{rec}, p_t^{ad}, c_t, rec_t)$ 
		\STATE  Execute action $a_{t}$ following off-policy $b(s_t)$ 
		\STATE  Calculate reward $r_t=r_t^{ad} + \alpha r_t^{ex}$ from offline log
		\STATE  Update state $s_t$ to $s_{t+1}$ 
		\STATE  Store transition $(s_{t}, a_{t},  r_{t}, s_{t+1})$ into $\mathcal{D}$
		\STATE  Sample mini-batch of transitions $(s, a, r, s')$ from $\mathcal{D}$
		\STATE  Set ${y=}
		\left\{\begin{array}{ll}
		r & \mathrm{terminal\,}  s'\\
		r{+}\gamma\max_{a'}Q(s'{,}a'{;}\theta) & \mathrm{non{-}terminal\,} s'
		\end{array}\right.$
		\STATE  Minimize  $\big(y-Q(s, a;\theta)\big)^{2}$ according to Eq.(\ref{equ:differentiating})
		\ENDFOR
		\ENDFOR
	\end{algorithmic}
\end{algorithm}

\begin{algorithm}[t]
	\small	
	\caption{\label{alg:model2} Online Test of the DEAR Framework.}
	\raggedright
	\begin{algorithmic} [1]
		\STATE Initialize the proposed DQN with well trained weights
		\FOR{session $=1, M$}
		\STATE  Initialize state $s_{0}$
		\FOR{$t=1, T$}
		\STATE  Observe state $s_t=concat(p_t^{rec}, p_t^{ad}, c_t, rec_t)$ 
		\STATE  Execute action $a_{t}$ following $Q^*(s_t,a_t)$
		\STATE  Observe rewards $r_t(s_t, a_t)$ from user
		\STATE  Update the state from $s_t$ to $s_{t+1}$ 
		\ENDFOR
		\ENDFOR
	\end{algorithmic}
\end{algorithm}

\subsection{The Optimization Task}
\label{sec:optimization}
The Deep Q-network, i.e., action-value function $Q(s_t, a_t)$, can be optimized by minimizing a sequence of loss functions $L(\theta)$ as:
\begin{equation}\label{equ:L}
L(\theta)=\mathbb{E}_{s_t, a_t,r_t,s_{t+1}}\big(y_t-Q(s_t, a_t;\theta)\big)^{2}
\end{equation}
\noindent where $y_t= \mathbb{E}_{s_{t+1}}[r_t+\gamma\max_{a_{t+1}}Q(s_{t+1},\ a_{t+1};\theta^{T})|s_t, a_t]$ is the target for the current iteration. We introduce separated evaluation and target networks~\cite{mnih2013playing} to help smooth the learning and avoid the divergence of parameters, where $\theta$ represents all parameters of the evaluation network, and the parameters of the target network $\theta^{T}$ are fixed when optimizing the loss function $L(\theta)$. The derivatives of loss function $L(\theta)$ with respective to parameters $\theta$:
\begin{equation}\label{equ:differentiating}
\nabla_{\theta}L(\theta) =\mathbb{E}_{s_t, a_t,r_t,s_{t+1}}\big(y_t{-}Q(s_t, a_t;\theta)\big)\nabla_{\theta}Q(s_t, a_t{;}\theta)
\end{equation}
\noindent where $y_t= \mathbb{E}_{s_{t+1}}[r_t+\gamma\max_{a_{t+1}}Q(s_{t+1},\ a_{t+1};\theta^{T})|s_t, a_t]$, and $\max_{a_{t+1}}$ will look through the candidate ad set $\{a_{t+1}^{ad}\}$ and all locations $\{a_{t+1}^{loc}\}$ (including the location that represents not interpolating an ad). Note that a recall mechanism is employed by the platform to select a subset of ads that may generate maximal revenue, and filter out ads that run out of their budget (RTB) or have fulfilled the guaranteed delivery amount (GD). In this paper, we mainly focus on the income of platform and user experience.

\subsection{Off-policy Training Task}
\label{sec:offline}
We train the proposed framework based on users' offline log, which records the interaction history between behavior policy $b(s_t)$ (the advertising strategy in use) and users' feedback. Our AA takes action based on the off-policy $b(s_t)$ and obtains the feedback from the offline log. We present our off-policy training algorithm in detail in Algorithm \ref{alg:model1}.

In each iteration of a training session, there are two stages. For storing transitions stage: given the state $s_t$ (line 6), the AA takes action $a_t$ according to the behavior policy $b(s_t)$ (line 7), which follows a standard off-policy way~\cite{degris2012off}; then the AA observes the reward $r_t$ from offline log (line 8) and updates the state to $s_{t+1}$ (line 9); and finally the AA stores transition $(s_t,a_t,r_t,s_{t+1})$ into replay buffer $\mathcal{D}$ (line 10). For model training stage: the AA samples minibatch of transitions $(s, a, r, s')$ from replay buffer $\mathcal{D}$ (line 11), and then updates the parameters according to Equation (\ref{equ:differentiating}) (lines 13). Note that in line 7, when the behavior policy $b(s_t)$ decides not to interpolate an ad, we use an all-zero vector as $a_{t}^{ad}$.

\subsection{Online Test Task}
\label{sec:on_testing}
The online test algorithm is presented in Algorithm~\ref{alg:model2}, which is similar to the transition generating stage in Algorithm~\ref{alg:model1}.  In each iteration of the session, given the current state $s_t$ (line 5), the AA decides to interpolate an ad into the rec-list (or not) by the well-trained advertising policy $Q^*(s_t,a_t)$ (line 6), then the target user browses the mixed rec-ad list and provides her/his feedback (line 7). Finally, the AA updates the state to $s_{t+1}$ (line 8) and goes to the next iteration.
\section{Experiments}
\label{sec:experiments}
In this section, we conduct extensive experiments on a real short video site to evaluate the effectiveness of the proposed framework. We mainly focus on three questions: (i) how the DEAR performs compared to representative baselines; (ii) how the components in the framework contribute to the performance; and (iii) how the hyper-parameters impact the performance. We first introduce experimental settings. Then we seek answers to these questions. 

\begin{table}
	\centering
	\label{table:statistics}
	\begin{tabular}{@{}|c|c|c|c|@{}}
		\toprule[1pt]
		\textbf{session}                                                & \textbf{user}                                                     & \textbf{normal video}                                                 & \textbf{ad video}                                                   \\
		{\small 1,000,000 }                                             & {\small 188,409}                                                  & {\small 17,820,066}                                                   & {\small 10,806,778 }                                                \\ \midrule
		\begin{tabular}[c]{@{}c@{}}\textbf{session}\\ \textbf{time}\end{tabular} & \begin{tabular}[c]{@{}c@{}}\textbf{session}\\ \textbf{length}\end{tabular} & \begin{tabular}[c]{@{}c@{}}\textbf{session}\\ \textbf{ad revenue}\end{tabular} & \begin{tabular}[c]{@{}c@{}}\textbf{rec-list}\\ \textbf{with ad}\end{tabular} \\
		{\small $\sim$ 18 min}                                             & {\small $\sim$ 55 videos }                                           & {\small 0.667}                                                        & {\small 55.23\%  }                                                  \\ \bottomrule[1pt]
	\end{tabular}
	\caption{Statistics of the Douyin video dataset.}
\end{table}

\subsection{Dataset}
\label{sec:Dataset}
We train our model on the dataset of March 1-30, 2019 collected in a short video site Douyin. 
There are two types of videos, i.e., normal videos (recommended items) and ad videos (advertised items). The features for a normal video contain: id, like score, finish score, comment score, follow score and group score, where the scores are predicted by the platform. The features for an ad video consist of: id, image size, bid-price, hidden-cost, predicted-ctr and predicted-recall, where the last four are predicted by the platform. Note that (i) the predicted features are successfully used in many applications such as recommendation and advertising in the platform, (ii) we discretize each feature as a one-hot vector, and (iii) the same features are used by baselines for a fair comparison. We collect 1,000,000 sessions in temporal order, and use first 70\% as training/validation set and other 30\% as test set. The statistics of the dataset are in Table \ref{table:statistics}. 

\subsection{Implementation Details}
\label{sec:Implement}
The dimensions of $p_t^{rec}$, $p_t^{ad}$, $c_t$, $rec_t$, $a_t^{ad}$ are 64, 64, 13, 360, 60. We leverage two 2-layer neural network to generate $V(s_t)$ and $A(s_t, a_t^{ad})$, respectively. The length of the output layer is $L+2=8$, i.e., there are $8$ possible locations, including the one representing not to interpolate an ad. We set the discounted factor $\gamma = 0.95$, and the replay buffer size is 10,000. For the hyper-parameters of the proposed framework such as $\alpha$, we select them via cross-validation. Correspondingly, we also do parameter-tuning for baselines for a fair comparison. We will discuss more details about hyper-parameter selection for the DEAR framework in the following subsections. Reward $r_t^{ad}$ is the revenue of ad videos, and $r_t^{ex}$ is 1 if user continue to browse next list and 0 otherwise.

\subsection{Metrics}
\label{sec:Metrics}
To measure the online performance, we leverage the \textit{accumulated rewards} in the session $R = \sum_{1}^{T} r_t$ as the metric, where $r_t$ is defined in Equation~(\ref{equ:reward}). To describe the significance of the results, we also use the \textit{improvement} of DEAR over each baseline and corresponding \textit{p$-$value}~\cite{ioannidis2018proposal} as metrics. Note that our task is to determine whether, which and where to insert an ad into a given rec-list, and our goal is to optimize long-term user experience and advertising revenue, which is different from ad click prediction task.

\begin{figure}[t]
	\centering
	\includegraphics[width=81mm]{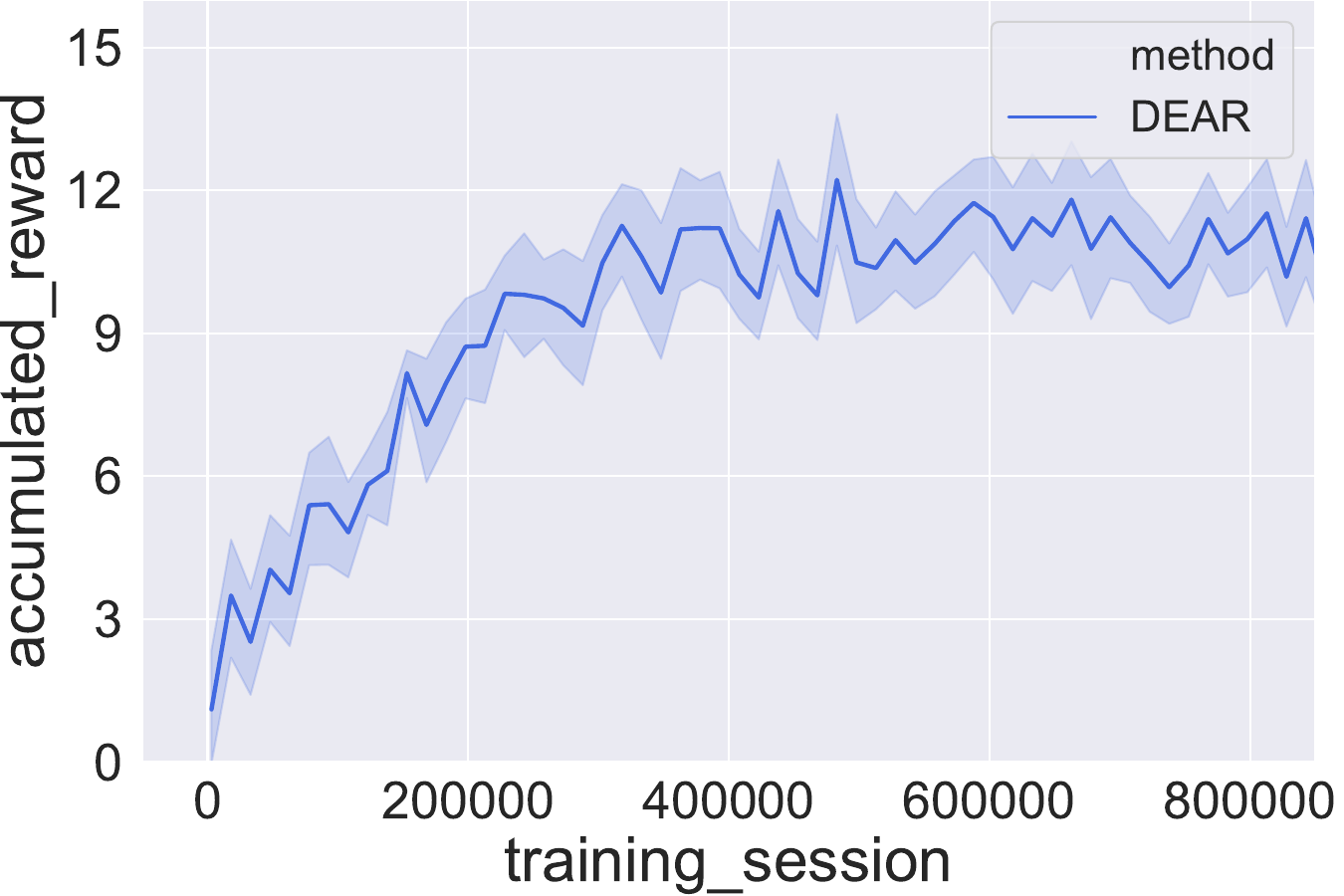}
	\label{fig:Fig5_result1}
	\caption{Model performance with training data.}
\end{figure}

\subsection{Baselines}
\label{sec:Baselines}
The experiment is based on a simulated online environment, which can simulate the rewards $r_t^{ad}$ and $r_t^{ex}$ according to a given state-action pair and a location. 
We compare the proposed framework with the following representative baseline methods: 
\textbf{W\&D} \cite{cheng2016wide}: This baseline is a wide \& deep model for jointly training feed-forward neural networks with embeddings and linear model with feature transformations for generic recommender systems with sparse inputs. We further augment its output layer to predict whether interpolate an ad and estimate the CTR of ads. W\&D is the behavior policy $b(s_t)$ in use of the video platform; 
\textbf{DFM} \cite{guo2017deepfm}: DeepFM is a deep neural network model that integrates the architectures of factorization-machine (FM) and wide \& deep model. It models low-order feature interactions like FM and models high-order feature interactions like W\&D; 
\textbf{GRU} \cite{hidasi2015session}: GRU4Rec utilizes RNN with Gated Recurrent Units (GRU) to predict what users will click/order next based on the clicking/ordering histories. We also augment its output layer for ads interpolation; 
\textbf{HDQN} \cite{kulkarni2016hierarchical}: This baseline is a hierarchical DQN framework where the high-level DQN determines the locations, and the low-level DQN selects a specific ad.
\begin{table}
	\centering
	\label{table:result1}
	\begin{tabular}{@{}|c|c|c|c|@{}}
		\toprule[1pt]
		method & reward & improvement & \textit{p$-$value} \\ \midrule
		W\&D   & 9.12   & 20.17\%     & 0.000     \\
		DFM    & 9.23   & 18.75\%     & 0.000     \\
		GRU    & 9.87   & 11.05\%     & 0.000     \\
		HDQN   & 10.27  & 6.712\%     & 0.002     \\
		\textbf{DEAR}   & \textbf{10.96}  & \textbf{-}           & \textbf{-}         \\ \bottomrule[1pt]
	\end{tabular}
	\caption{Overall performance comparison.}
\end{table}

\subsection{Overall Performance Comparison}
\label{sec:ev_overall}
The overall performances are shown in Table~\ref{table:result1}. We make the following observations:
Figure~\ref{fig:Fig5_result1} illustrates the training process of the proposed model. It can be observed that Read converges with around 600,000 training sessions; 
The DFM achieves better performance than W\&D, where DeepFM can be trained end-to-end without any feature engineering, and its wide part and deep part share the same input and also the embedding vector;
GRU outperforms W\&D and DFM, since GRU can capture the temporal sequence of user behaviors within one session, while W\&D and DFM neglect it;
HDQN performs better than GRU, since GRU is designed to maximize each request's immediate reward, while HDQN aims to maximize the rewards in the long run. This result suggests that introducing reinforcement learning can improve the long-term performance of online recommendation and advertising;
Our proposed model DEAR outperforms HDQN, since HRL frameworks like HDQN are not stable when multiple levels are jointly trained in an off-policy manner~\cite{nachum2018data}.                                
To sum up, DEAR outperforms representative baselines with significant margin (\textit{p$-$value $<$ 0.01})~\cite{ioannidis2018proposal}, which demonstrates its effectiveness in online advertising. 

\subsection{Component Study}
\label{sec:component}
To answer the second question, we systematically eliminate the corresponding components of DEAR by defining the following variants:
\textbf{DEAR-1}: This variant shares the same architectures with the proposed model while training the framework through a supervised learning manner.
\textbf{DEAR-2}: This variant is to evaluate the effectiveness of RNNs, hence we replace each RNN with two fully-connected layers (FCNs), concatenate recommended or advertised items as one vector and feed it into the corresponding FCN.
\textbf{DEAR-3}: This baseline leverages the DQN architecture in Figure~\ref{fig:Fig2_DQNs}(b) with an additional input, which represents the location by a one-hot vector.
\textbf{DEAR-4}: The architecture of this variant does not divide the Q-function into the value function $V(s)$ and the advantage function $A(s, a)$ for the AA.
\textbf{DEAR-5}: In this variant, we replace the selected ad with a random ad from all candidate ads.
\textbf{DEAR-6}: In this variant, we insert the selected ad into a random slot.

The results are shown in Table \ref{table:result2}. It can be observed:
DEAR-1 validates the effectiveness of introducing reinforcement learning for online advertising.
DEAR-2 demonstrates that capture user's sequential preferences over recommended and advertised items can boost performance.
DEAR-3 validates the effectiveness of the proposed DEAR architecture over conventional DQN architecture that takes an ad $a_t^{ad}$ as input while outputs the Q-value corresponding to all possible locations $\{a_t^{loc}\}$ for the given ad $a_t^{ad}$.
DEAR-4 proves that whether interpolate an ad into rec-list is mainly dependent on state (especially the quality of rec-list), while the reward for selecting an ad and location depends on both $s_t$ and $a_t$ (ad). Thus dividing $Q(s_t, a_t)$ into the value function $V(s_t)$ and the advantage function $A(s_t, a_t)$ can improve the performance.
DEAR-5 and DEAR-6 demonstrate that the ads and slot influence the performance, and DEAR indeed selects a proper ad and inserts it into a proper slot.
In summary, introducing RL and appropriately designing neural network architecture can boost performance.

\begin{table}
	\centering
	\label{table:result2}
	\begin{tabular}{@{}|c|c|c|c|@{}}
		\toprule[1pt]
		variant & reward         & improvement & \textit{p$-$value} \\ \midrule
		DEAR-1  & 9.936          & 10.32\%     & 0.000              \\
		DEAR-2  & 10.02          & 9.056\%     & 0.000              \\
		DEAR-3  & 10.39          & 5.495\%     & 0.001              \\
		DEAR-4  & 10.57          & 3.689\%     & 0.006              \\
		DEAR-5  & 9.735          & 12.58\%     & 0.000              \\
		DEAR-6  & 9.963          & 10.01\%     & 0.000              \\
		\textbf{DEAR}    & \textbf{10.96} & \textbf{-}  & \textbf{-}         \\ \bottomrule[1pt]
	\end{tabular}
	\caption{Component study results.}
\end{table}

\subsection{Parameter Sensitivity Analysis}
\label{sec:parametric}
\begin{figure}[t]
	\centering
	\hspace*{-14.9mm}{\subfigure{\includegraphics[width=0.3856\linewidth]{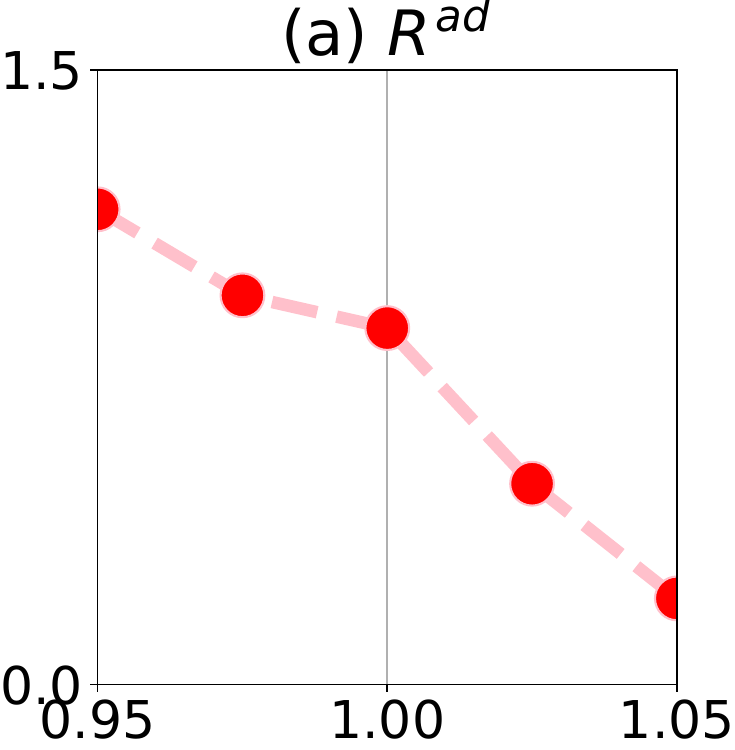}}}
	{\subfigure{\includegraphics[width=0.3856\linewidth]{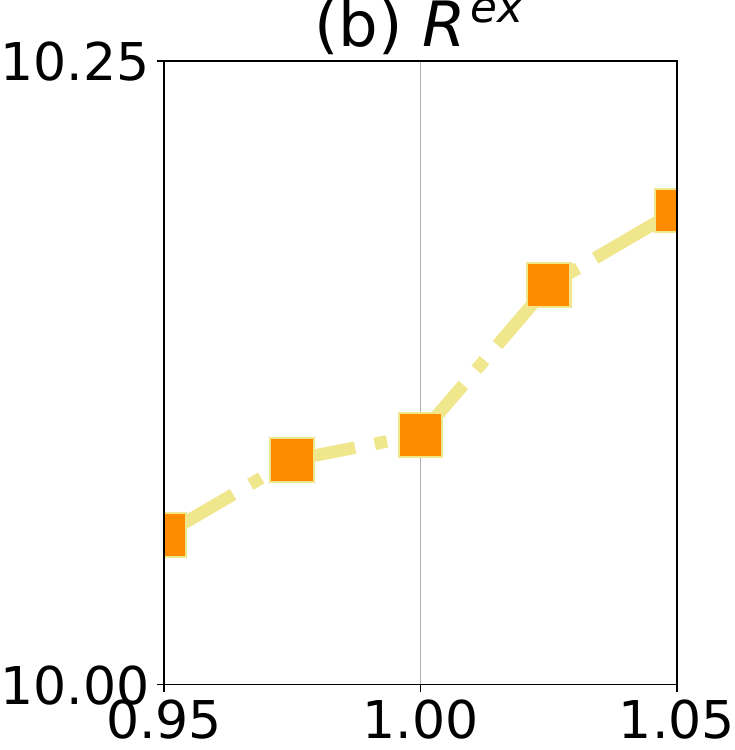}}}
	{\subfigure{\includegraphics[width=0.3856\linewidth]{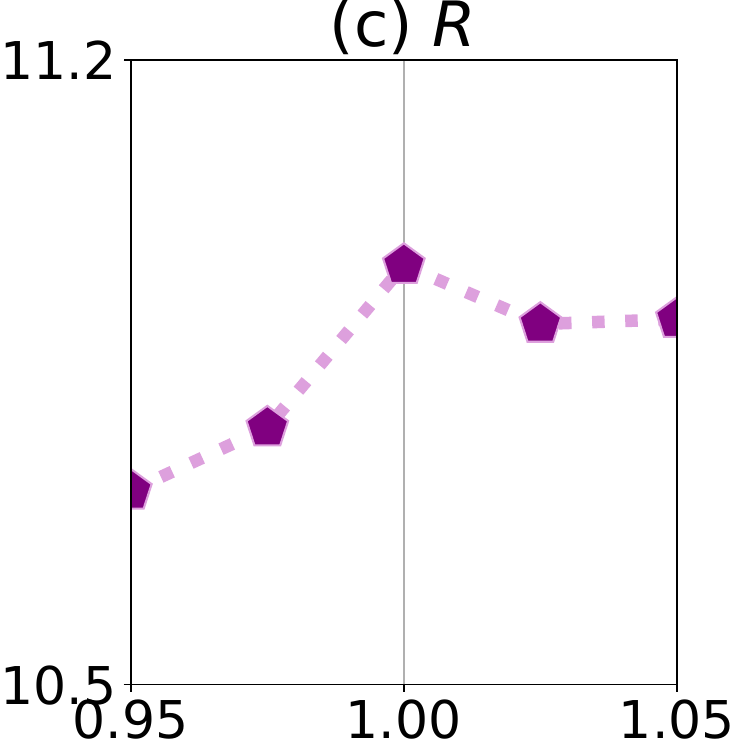}}}
	\caption{Parameter sensitivity analysis.}
	\label{fig:Fig5_result3}
\end{figure}

In this section, we investigate how the proposed framework DEAR performs with the changes of $\alpha$ in Equation~(\ref{equ:reward}), while fixing other parameters. We select the accumulated rewards $R^{ad}=\sum_{1}^{T} r_t^{ad}$, $R^{ex}=\sum_{1}^{T} r_t^{ex}$, and $R=R^{ad} + \alpha R^{ex}$ of the whole session as the metrics. 
In Figure~\ref{fig:Fig5_result3} (a)-(b) we find that with the increase of $\alpha$, $R^{ex}$ improves while $R^{ad}$ decreases. On the one hand, when we increase the importance of the second term in Equation~(\ref{equ:reward}), the AA tends to insert fewer ads or select the ads that will not decrease user's experience, although they may generate suboptimal revenue. On the other hand, when we decrease the importance of the second term, the AA prefers to insert more ads or choose the ads that will lead to maximal revenue, while ignoring the negative impact of ads on user's experience. Therefore, online platforms should carefully select $\alpha$ according to their business demands (in our setting, the overall $R$ achieves the peak at $\alpha=1$ as shown in Figure~\ref{fig:Fig5_result3} (c)). 
\section{Related Work}
\label{sec:related_work}
In this section, we briefly review works related to our study. In general, the related work can be mainly grouped into the following categories.

The first category related to this paper is RL-based online advertising, including guaranteed delivery (GD) and real-time bidding (RTB). GD charges ads on a pay-per-campaign basis for the pre-specified number of deliveries~\cite{salomatin2012unified}. In~\cite{wu2018multi}, a multi-agent reinforcement learning~(MARL) approach is proposed to derive cooperative policies for the publisher to maximize its target in an unstable environment. RTB allows an advertiser to submit a bid for each impression in a very short time frame under the multi-armed bandits setting~\cite{yang2016dynamic,nuara2018combinatorial,gasparini2018targeting,tang2013automatic,xu2013estimation,yuan2013adaptive,schwartz2017customer}. 
Thus, the MDP setting has also been studied to tackle that a given ad campaign would repeatedly happen during its life span before the budget running out~\cite{cai2017real,wang2018learning,rohde2018recogym,wu2018budget,jin2018real}.

%

The second category related to this paper is RL-based recommender systems~\cite{zhao2019deep,zhang2020deep}. 
Users' positive and negative feedback, i.e., purchase/click and skip behaviors, are jointly considered in one framework to boost recommendations, since both types of feedback can represent part of users' preference~\cite{zhao2018recommendations}. 
A page-wise recommendation framework is proposed to jointly recommend a page of items and display them within a 2-D page~\cite{zhao2017deep,zhao2018deep}. 
A multi-agent model-based reinforcement learning framework (DeepChain) is proposed for the whole-chain recommendation problem~\cite{zhao2020whole}.
A user simulator RecSimu base on Generative Adversarial Network (GAN) framework is presented for RL-based recommender systems~\cite{zhao2019toward}.
Other related work in this category include~\cite{zhao2020jointly,fan2020attacking,zou2020neural,dulac2015deep,chen2018large,choi2018reinforcement,zheng2018drn,wang2018reinforcement,chen2018top,liu2020automated,ge2021towards}
\section{Conclusion}
\label{sec:conclusion}
In this paper, we propose a deep reinforcement learning framework DEAR with a novel Deep Q-network architecture for online advertising in recommender systems. It is able to (i) determine three internally related actions at the same time, i.e., whether to interpolate an ad in a rec-list or not, if yes, which is the optimal ad and location to interpolate; and (ii) simultaneously maximize the revenue of ads and minimize the negative influence of ads on user experience. It is worth noting that the proposed DQN architecture can take advantage of two conventional DQN architectures, which can evaluate the Q-value of two or more kinds of related actions simultaneously. We evaluate our framework with extensive experiments based on a short video site. The results show that our framework can significantly improve online advertising performance in recommender systems.

\section{Acknowledgments}
This work is supported by National Science Foundation (NSF) under grant numbers IIS1907704, IIS1928278, IIS1714741, IIS1715940, IIS1845081 and CNS1815636.

\bibliographystyle{aaai21}
\bibliography{9Reference}

\end{document}